\documentclass[a4paper]{jpconf}

\usepackage{graphicx}

\usepackage{iftex}
\ifPDFTeX
  \usepackage[UKenglish]{babel}

  \usepackage[utf8]{inputenc}
  \usepackage[T1]{fontenc}
  \usepackage{lmodern}
\else	
  \usepackage{polyglossia}
  \setdefaultlanguage[variant=uk]{english}

  \usepackage{fontspec}
  \defaultfontfeatures{Ligatures=TeX}
\fi

\graphicspath{{img/}{../../images/}}

\begin{document}

\title{RootJS: Node.js Bindings for ROOT~6}
\author{Theo~Beffart, Maximilian~Früh, Christoph~Haas, Sachin~Rajgopal, Jonas~Schwabe, Christoph~Wolff, Marek~Szuba}
\address{Steinbuch Centre for Computing, Karlsruhe Institute of Technology, Hermann-von-Helmholtz-Platz~1, 76344~Eggenstein-Leopoldshafen, Germany}
\ead{Marek.Szuba@cern.ch}

\begin{abstract}
  We present rootJS, an interface making it possible to seamlessly
  integrate ROOT~6 into applications written for Node.js, the
  JavaScript runtime platform increasingly commonly used to create
  high-performance Web applications. ROOT features can be called both
  directly from Node.js code and by JIT-compiling C++ macros. All
  rootJS methods are invoked asynchronously and support callback
  functions, allowing non-blocking operation of Node.js applications
  using them. Last but not least, our bindings have been designed to
  platform-independent and should therefore work on all systems
  supporting both ROOT 6 and Node.js.

  Thanks to rootJS it is now possible to create ROOT-aware Web
  applications taking full advantage of the high performance and
  extensive capabilities of Node.js. Examples include platforms for
  the quality assurance of acquired, reconstructed or simulated data,
  book-keeping and e-log systems, and even Web browser-based data
  visualisation and analysis.
\end{abstract}

\section{Introduction}

\subsection{Background}

The scientific software framework ROOT is a \textit{de facto} standard
tool in nuclear and particle physics~\cite{Brun:1997pa}. It provides
components for a wide range of purposes such as statistical data
analysis, advanced visualisation, machine learning, Monte-Carlo
simulations, persistence of data, parallel processing, and many
more. Although it has been written mainly in C++ and typically
interfaced with using macros (scripts) written in that language, ROOT
bindings exist for other languages such as Python, R, or Ruby.

Among its numerous capabilities ROOT features a set of building blocks
for graphical user interfaces. ROOT-based GUI applications include
event viewers, control and monitoring applications, and data
quality-assurance tools. However, such tools have traditionally been
built as monolithic applications which require ROOT and all of its
dependencies installed locally and, unless carefully designed for that
purpose, offered minimal scalability.

A well-known alternative to monolithic tools are Web applications ---
client-server software in which the client runs in a Web
browser. Unlike traditional software, Web applications naturally
follow a distributed model of multiple tiers, each of them responsible
for a different part of application logic. This model greatly improves
scalability because different tiers can be extended
independently. Moreover not only is the client component of a Web
application typically lightweight, it also only requires the user to
run a modern, standard-compliant Web browser --- doing away with the
need for local installation or creating builds for many different
operating systems and hardware architectures. At the same time, modern
Web standards and techniques greatly reduce constraints imposed by
having to run an application inside a Web browser. To name just three:
Ajax allows for the the user interface to be dynamically updated
inside the browser without the need for reloading the entire page,
WebGL leverages local GPU power to render 3D graphics, and IndexedDB
enables the use transactional local databases in applications.

\subsection{Node.js and the MEAN Stack}

One of the defining characteristics of a Web application is the
\emph{software stack} it is based on. Possibly the best known among
these is LAMP, which originally consisted of the Linux operating
system, the Apache HTTP Server, the MySQL relational database
management system and the PHP programming language, and which along
with its derivatives remains arguably the most popular model on the
Internet~\cite{Kunze:1998aa}.

An alternative, more recent Web-application stack is known as MEAN and
consists of the NoSQL database MongoDB, the Web-application framework
Express running on top of the server-side platform Node.js, and the
front-end framework AngularJS~\cite{web:MEANstack}. It offers
several advantages to both users and developers. For example: its
components have been specifically designed for high performance, it
naturally allows the client to perform more tasks than the mere
rendering of content fetched from the server, and the use of a single
programming language throughout the stack (JavaScript) simplifies both
development and debugging.

Of the four MEAN components, the one directly relevant to the present
paper is Node.js. It is an open-source and cross-platform JavaScript
runtime environment based on the V8 JavaScript engine created by
Google. The core runtime is minimalistic by design but can be extended
with modules (for instance one to provide a HTTP server, which is how
one most commonly uses Node in the server-side tier of Web
applications), the management of which is simplified by the Node
Package Manager (\texttt{npm}) and the corresponding public module
directory~\cite{web:npmjs}. An important feature of Node.js is its
high performance, which it owes to the fact that it is almost entirely
asynchronous --- instead of blocking on I/O operations, it handles
their results via callback functions.

\subsection{Motivation}

The motivation behind work described in the present paper has been to
develop Open Source JavaScript bindings for ROOT, which we have called
\textit{rootJS}, which would allow it to be used in Node.js scripts
and MEAN stack-based Web applications. The following sections of the
manuscript discuss the architecture of these bindings, outlines some
of the details of their implementation along with associated
challenges, and discusses potential use cases.  Finally, we discuss
related work.

\section{Overview of rootJS}

\subsection{Functional Requirements}

The most important requirement we defined for our bindings was that
they must be complete. Given the ROOT class library already numbers
into thousands, is continuously extended by upstream developers, and
can be supplemented at both compile and run time by third-party
components such as those created by specific experiments, this
requires the lists of classes and methods available through rootJS to
be handled dynamically. On a related note, we also wanted to allow the
users to load additional ROOT libraries at run time.

Secondly, the bindings must take the form of a Node.js module
\textit{i.e.} require no modifications to Node.js source
code. Similarly they should communicate with ROOT using its standard
APIs, again not requiring modifications of upstream code.

Next, the bindings should keep with the spirit of Node.js and
provide asynchronous wrappers for at least the most common I/O
functions. Ideally they should support asynchronous execution of all
ROOT functions.

Finally, the bindings should allow the user to feed C++ code to the
ROOT just-in-time-compiler to support re-use of existing code
blocks. Needless to say rootJS must accurately reflect the state of
ROOT following execution of such code, which provided another argument
in favour of the dynamic approach.

\subsection{Software Requirements}

We have decided rootJS would only support ROOT~6 because its Low Level
Virtual Machine (LLVM)-based C++ interpreter Cling offers many
advantages, \textit{e.g.} the aforementioned just-in-time compilation,
over the one available in older ROOT versions.

Regarding operating-system compatibility, our goal has been for the
bindings to support GNU/Linux on amd64 architecture. That said,
completed bindings have been shown to run correctly under Mac OS X and
should in theory be compatible with all systems supported by both
ROOT~6 and Node.js.

\subsection{Challenges}

Developing JavaScript bindings for a C++ library like ROOT is not a
trivial task because of fundamental differences between the two
languages. Most notably:
\begin{itemize}
\item JavaScript is a functional language with first-class functions;
  the latter feature is widely used for callbacks. In contrast, C++ is
  an imperative and object-oriented language in which functions are
  not first-class citizens;
\item they employ different type systems: C++ is strongly typed,
  JavaScript offers only limited dynamic type checking. One important
  consequence of the above is that a JavaScript engine will not
  overload functions depending on the type of its arguments, which is
  widely applied in ROOT (see \textit{e.g.}
  \texttt{TH1::Fill(Double\_t, Double\_t)} \textit{vs}
  \texttt{TH1::Fill(const~char*, Double\_t)});
\item on a related note, the two languages support different data
  types. For instance: JavaScript only has a single numeric type; only
  C++ supports enumerated types; strings are a primary data type in
  JavaScript but objects in C++;
\item C++ supports low-level memory access and can require the user to
  implement their own memory management. JavaScript has neither;
\item unlike C++, JavaScript objects are classless.
\end{itemize}

Fortunately both Node.js and ROOT provide ways in which these
differences can be reconciled in order to create an \emph{adapter} between the two:
\begin{itemize}
\item the C++ API of the V8 engine allows exposure of C++ objects to
  Node.js, can map JavaScript objects to C++ classes, and provides
  callback handling;
\item ROOT features extensive run-time introspection capabilities,
  covering both individual functions/objects as well as general
  information about classes, name spaces, and global and member
  variables.
\end{itemize}

\section{Architecture}

The following elements have been implemented in rootJS in order for it
to meet our requirements.

To begin with, the bindings recursively seek and expose ROOT classes
and name spaces. This is done at initialisation time, after the
loading of a new library, after execution of JIT-compile code, or upon
user request. The name-space hierarchy of ROOT is preserved and
reproduced by a tree of properties attached to the top-level object of
the bindings.

Secondly, existing objects, functions and variables must be wrapped in
appropriate \textit{proxy objects} which are then exposed by V8 to
JavaScript; similar encapsulation is also established for newly
created ROOT objects after the forwarding of respective constructor
calls. These proxy objects are produced by respective
\emph{factories}, which also cache their products so that duplication
of work is avoided \textit{e.g.} when multiple instances of the same
ROOT object are created. The proxies also normalise memory addresses
in order to properly handle the use of pointers common in ROOT.

It is worth emphasising at this point that all read and write
operations on exposed objects and variables take place directly in
ROOT memory space. That way everything stays in sync all the time.

Furthermore, function calls must be provided asynchronous call context
in order to avoid blocking. We have implemented concurrent execution
of functions using \textit{libuv}, an asynchronous-I/O library
designed for Node.js for this very purpose~\cite{web:libuv}; libuv
maintains a pool of worker threads and communicates with them using
messages. The original idea to employ the threading API provided by
ROOT (\texttt{TThread}) had to be abandoned due to limitations of the
V8 engine, which requires all interaction with Node.js to take place
in the main thread.

Finally, a handler exists which maintains associations between
function calls and their respective callbacks.

\begin{figure}
  \centering
  \includegraphics[width=\columnwidth]{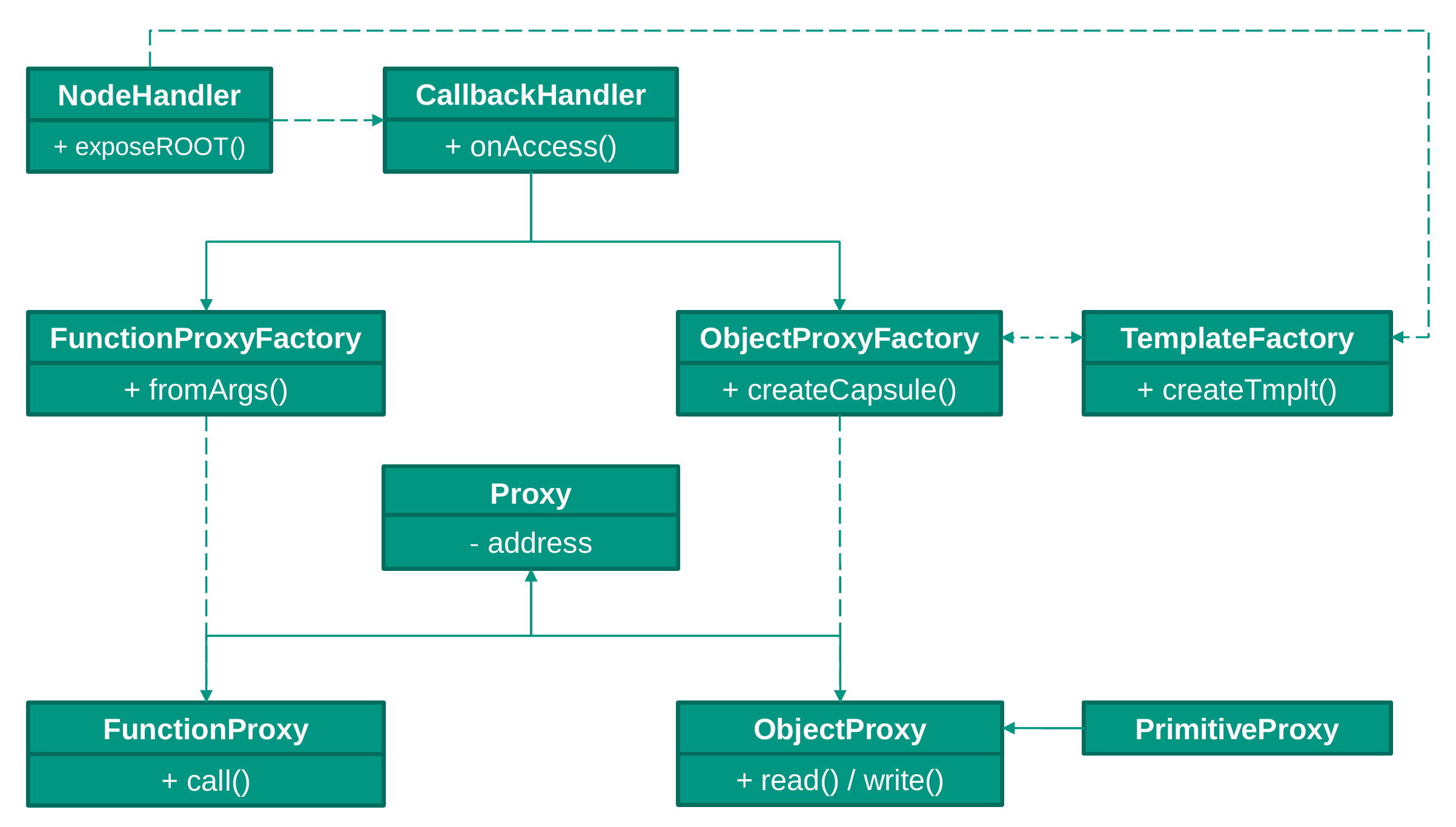}
  \caption{Core architecture of rootJS.}
  \label{fig:coreArchitecture}
\end{figure}

A diagram of the core architecture of rootJS can be found in
Figure~\ref{fig:coreArchitecture}.

\section{Getting and Using rootJS}

rootJS has been registered in the npm module directory so for most
users who have already got Node.js and its package manager installed,
getting rootJS grinds down to a single command:

\begin{verbatim}
npm install rootjs
\end{verbatim}

Alternatively one can manually download the source code from our
GitHub repository or its CERN GitLab mirror~\cite{web:rootJSGitHub,
  web:rootJSCERN}.

Please note that rootJS is a native C++ module that has to be compiled
before use, which requires a working C++ compiler as well as libuv
header files. Naturally one must also have ROOT~6 installed.

After installation the module is loaded the standard Node.js way:

\begin{verbatim}
var root = require('rootjs');
\end{verbatim}

after which all ROOT variables, name spaces and classes become
available through the object \texttt{root}.

rootJS is ultimately a thin wrapper for ROOT itself so most usage
examples would be the same (including operations producing graphics on
the screen --- at least under Linux, using rootJS does not prevent
ROOT from communicating with the local X server), making it
unnecessary to repeat them here. An exception to this are callbacks,
which can always be passed to functions as their last argument. For
example:

\begin{verbatim}
root.TFile.Open("foo.root", function (fin) {
  fin.ls();
});
\end{verbatim}

will return immediately, then only invoke \texttt{ls()} on the file
once it has been opened. One can of course still invoke functions
without callbacks in which case they are executed the standard ROOT
way, only returning once the call has been finished.

Finally, the loading of additional libraries works slightly
differently so that rootJS can immediately update its lists of classes
and variables. Instead of using \texttt{gSystem->Load()}, one does the
following:

\begin{verbatim}
root.loadlibrary("libMathCore.so"); // load it...
root.ROOT.Math.Pi(); // ...and use it! Note the preservation of name spaces
\end{verbatim}

For additional information, please see our GitHub
site~\cite{web:rootJSGitHub}.

\section{Use Cases}

Like ROOT itself, rootJS is a set of building blocks to create various
applications rather than an actual application. Its direct purpose is
simply to bring the capabilities of ROOT to another programming
language (JavaScript), with emphasis on a particular domain
(high-performance Web applications). That said, we would like to
discuss at this point two possible use cases for rootJS which have
inspired us to initiate this project.

On one hand, rootJS could be used to implement a Web-based event
viewer for a HEP experiment. Such viewers are commonly used for the
monitoring of data taking by experiments as well as basic quality
assurance, and they are not infrequently implemented using ROOT. There
are, however, two disadvantages of using an event viewer in the form
of a standalone application:
\begin{itemize}
\item limited portability --- it must be installed on the target
  machine, along with ROOT and all other dependencies. This is at
  present an issue especially for applications requiring ROOT~6 due to
  its still-limited availability (\textit{e.g.} lack of native Windows
  versions);
\item such a viewer is usually located close to the data. This is
  particularly important for live monitoring (in which case the source
  of data is typically isolated from the Internet) but can also be an
  issue in off-line mode (in which case the data might be too big to
  move around quickly).
\end{itemize}
Using a modern, multi-tiered Web application instead can improve the
situation: on one hand all the end user needs is a Web browser, on the
other they can be located essentially anywhere because only the
back-end tier has to be located near the data source. There has been
growing interest in Web-based event browsers in the particle-physics
community (see \textit{e.g.}~\cite{thisCHEP:McCauley} in these
proceedings) --- and rootJS enables direct integration of ROOT into
one of the increasingly popular back-end platforms.

On the other hand, rootJS makes it possible to employ the capabilities
of ROOT for virtually any sort of analytics in MEAN stack-based Web
applications. It could be used to integrate machine learning,
statistical analysis, linear algebra \dots but also access to XRootD,
PROOF and so on. In our particular case, we plan to take advantage of
ROOT in Web-based analysis of data from Earth-observing climatology
satellites~\cite{DBLP:conf/ccgrid/SzubaAGMS16}.

\section{Related Work}

In this section we would like to briefly introduce three examples of
related work --- JavaScript ROOT, ROOT \texttt{THttpServer}, and the ROOT
kernel for Jupyter Notebook --- and compare them to rootJS. These
examples have been chosen not only because they are relevant to the
use cases described above but also because they are now part of
the core ROOT distribution.

\subsection{JavaScript ROOT}

In the words of its authors, ``JavaScript ROOT provides interactive
ROOT-like graphics in the web browsers. Data can be read and displayed
from binary and JSON ROOT files. JSROOT implements user interface for
THttpServer class.'' It is a JavaScript reimplementation of the
aforementioned components of ROOT.

This description alone makes it clear that JSROOT and rootJS are
complementary rather than competing solutions --- JSROOT is a
\emph{front-end} component of a Web application whereas our bindings
would be used on the \emph{back-end} tier. It would be entirely
feasible to create an application in which users interact with the
former but in which all the under-the-bonnet processing is handled
through the latter.

\subsection{ROOT THttpServer}

As mentioned above, ROOT~6 contains an integrated HTTP
server. \texttt{THttpServer} is based on the embedded C/C++ server
CivetWeb and continues to be extended with additional
features~\cite{web:Civet, thisChep:Naumann}. A question might
therefore arise: why bother with Node.js and rootJS if the same can be
achieved with ROOT itself?

Leaving aside the subjects of performance (which would have to be
carefully benchmarked before any conclusions could be made) or
individual preferences, we believe the key features in favour of
rootJS are scalability and the wide range of modules available for
Node.js which extend its capabilities as a Web server. CivetWeb, and
by extension \texttt{THttpServer}, is a capable yet lightweight tool
which for a lot of applications will be more than enough. However,
what if you wanted to migrate your back-end server to HTTP2? Enable
U2F-compliant two-factor authentication? Move from RESTful HTTP to
WebSockets? Or perhaps just replace your single Web server with a
high-availability cluster? In case of Node.js, all of these features
are already supported --- in some cases as easily as by a drop-in
replacement of a module.

\subsection{Jupyter Notebook}

Jupyter Notebook is a Web application designed for creation and
sharing of documents containing live code (in over 40 languages),
equations, visualizations and explanatory
text~\cite{DBLP:conf/elpub/KluyverRPGBFKHG16}. A ROOT kernel has been
developed for Jupyter which enhances the latter with support for C++
ROOT code~\cite{thisChep:Piparo}. Moreover, by using the Python~2
kernel and the ROOT Python bindings (PyROOT) one interface ROOT with
the numerous data-science Python modules such as SciPy.

While there is a certain overlap between Jupyter Notebook and rootJS,
the two are considerably different: the former is a complete Web
application featuring a sophisticated interactive interface, whereas
the latter is a library interface aimed at scripted and batch
processing on the back-end tier. Ultimately, they cover radically
different use cases.

It is also worth noting at this point that Project Jupyter also
features a JavaScript kernel that internally uses Node.js. Using
rootJS with this kernel one can therefore grant ROOT notebooks access
to the Node.js module ecosystem.

\section{Conclusions}

We have developed cross-platform Node.js bindings for ROOT~6, called
rootJS, which makes it possible to integrate capabilities of ROOT into
high-performance Web applications based on the MEAN stack. The
bindings provide access to all classes and variables of ROOT,
including those provided by additional libraries loaded at run
time. They also allow the user to invoke all ROOT functions
asynchronously with callbacks.

\section*{Acknowledgements}

We would like to express our thanks to the authors of PyROOT, which
has proven an extremely helpful reference for the Cling API.

The bulk of this work was done as part of the Software Engineering
Practice class for Computer Science students of Karlsruhe Institute of
Technology.

This work was conducted within the scope of the project ``Large-Scale
Data Management and Analysis'', funded by the German Helmholtz
Association~\cite{Jung:2014oba}.

\section*{References}

\bibliographystyle{iopart-num}
\bibliography{references}

\providecommand{\newblock}{}
\begin{thebibliography}{10}
\expandafter\ifx\csname url\endcsname\relax
  \def\url#1{{\tt #1}}\fi
\expandafter\ifx\csname urlprefix\endcsname\relax\def\urlprefix{URL }\fi
\providecommand{\eprint}[2][]{\url{#2}}

\bibitem{Brun:1997pa}
Brun R and Rademakers F 1997 {\em Nucl. Instrum. Meth.\/} {\bf A389} 81--86

\bibitem{Kunze:1998aa}
Kunze M 1998 {\em c't Archiv\/}  230 (in German)

\bibitem{web:MEANstack}
Karpov V 2013 {The MEAN Stack: MongoDB, ExpressJS, AngularJS and Node.js} (last
  visited: 2017-02-20)
  \urlprefix\url{https://www.mongodb.com/blog/post/the-mean-stack-mongodb-expressjs-angularjs-and}

\bibitem{web:npmjs}
{node package manager} (last visited: 2017-02-20)
  \urlprefix\url{https://www.npmjs.com/}

\bibitem{web:libuv}
{libuv | Cross-platform asynchronous I/O} (last visited: 2017-02-20)
  \urlprefix\url{http://libuv.org/}

\bibitem{web:rootJSGitHub}
\urlprefix\url{https://github.com/rootjs/rootjs}

\bibitem{web:rootJSCERN}
\urlprefix\url{https://gitlab.cern.ch/rootjs/rootjs}

\bibitem{thisCHEP:McCauley}
Mc~Cauley T {A browser-based event display for the CMS Experiment at the LHC
  using WebGL} {\em these proceedings\/}

\bibitem{DBLP:conf/ccgrid/SzubaAGMS16}
Szuba M, Ameri P, Grabowski U, Meyer J and Streit A 2016 A distributed system
  for storing and processing data from earth-observing satellites: System
  design and performance evaluation of the visualisation tool {\em {IEEE/ACM}
  16th International Symposium on Cluster, Cloud and Grid Computing, CCGrid
  2016, Cartagena, Colombia, May 16-19, 2016\/} ({IEEE} Computer Society) pp
  169--174 ISBN 978-1-5090-2453-7
  \urlprefix\url{http://dx.doi.org/10.1109/CCGrid.2016.19}

\bibitem{web:Civet}
{CivetWeb: Embedded C/C++ web server} (last visited: 2017-02-20)
  \urlprefix\url{https://github.com/civetweb/civetweb}

\bibitem{thisChep:Naumann}
Naumann A {Status and Evolution of ROOT} {\em these proceedings\/}

\bibitem{DBLP:conf/elpub/KluyverRPGBFKHG16}
Kluyver T, Ragan{-}Kelley B, P{\'{e}}rez F, Granger B~E, Bussonnier M, Frederic
  J, Kelley K, Hamrick J~B, Grout J, Corlay S, Ivanov P, Avila D, Abdalla S,
  Willing C and et~al 2016 Jupyter notebooks - a publishing format for
  reproducible computational workflows {\em Positioning and Power in Academic
  Publishing: Players, Agents and Agendas, 20th International Conference on
  Electronic Publishing, G{\"{o}}ttingen, Germany, June 7-9, 2016.\/} ed
  Loizides F and Schmidt B ({IOS} Press) pp 87--90 ISBN 978-1-61499-648-4
  \urlprefix\url{http://dx.doi.org/10.3233/978-1-61499-649-1-87}

\bibitem{thisChep:Piparo}
Piparo D, Tejedor~Saavaedra E and Mato~Vila P {The New ROOT Interface: Jupyter
  Notebooks} {\em these proceedings\/}

\bibitem{Jung:2014oba}
Jung C, Gasthuber M, Giesler A, Hardt M, Meyer J, Rigoll F, Schwarz K, Stotzka
  R and Streit A 2014 {\em J. Phys. Conf. Ser.\/} {\bf 513} 032047

\end{thebibliography}

\end{document}